\documentclass[12 pt]{article}
\usepackage{graphicx}
\usepackage[utf8]{inputenc}
\usepackage{amsmath}
\usepackage{amssymb}
\usepackage{geometry}
\usepackage{lineno}
\usepackage{textcomp}
\usepackage{xcolor}
\usepackage{slashed}
\usepackage{dcolumn}
\usepackage{bm}
\usepackage{titlecaps}
\usepackage{authblk}
\usepackage[colorlinks,citecolor=blue]{hyperref}
\usepackage{tikz}
\usepackage[compat=1.1.0]{tikz-feynman}
\usepackage[sorting=none]{biblatex}
\begin{document}
	
\title{\bf Leptophilic-portal Dark Matter in the Light of AMS-02 positron excess}

\author[1]{{\bf Sayan Ghosh}\footnote{sayan.ghosh@saha.ac.in}}
\author[2]{{\bf Amit Dutta Banik}\footnote{ amitdbanik@mail.ccnu.edu.cn}}
\author[3]{{\\ \bf Eung Jin Chun}\footnote{ejchun@kias.re.kr}}
\author[1]{{\bf Debasish Majumdar}\footnote{debasish.majumdar@saha.ac.in}}
\affil[1]{\textit{Saha Insitute of Nuclear Physics, HBNI, 1/AF Bidhannagar,  
		Kolkata-700~064, India.}}
\affil[2]{\textit{Key Laboratory of Quark and Lepton Physics (MoE) and Institute of Particle Physics, Central China Normal University, Wuhan 430079, China.}}
\affil[3]{\textit{Korea Institute for Advanced Study, 85 Hoegiro, Seoul 02455, Republic of Korea.}}

\date{}

\maketitle

\begin{abstract}	
	We revisit dark matter annihilation as an explanation of the positron excess reported recently by the AMS-02 satellite-borne experiment. To this end, we propose a particle dark matter model by considering a Two Higgs Doublet Model (2HDM) extended with an additional singlet boson and a singlet fermion. The additional (light) boson mixes with the pseudoscalar inherent in the 2HDM, and the singlet fermion, which is the dark matter candidate, annihilates via this bosonic portal. The dark matter candidate is made leptophilic by choosing the lepton-specific 2HDM and a suitable high value of $\tan\beta$. We identify the model parameter space which explains the muon g-2 anomaly while evading the experimental constraints. After establishing the viability of the singlet fermion to be a dark matter candidate, we calculate the positron excess produced from its annihilation to the light bosons which primarily decay to muons. Incorporating the Sommerfeld effect caused by the light mediator and an appropriate boost factor, we find that our proposed model can satisfactorily explain the positron fraction excess as well as the positron spectrum data reported by AMS-02 experiment. 
\end{abstract}

\newpage
\section{Introduction}
The existence of Dark Matter in the Universe has now been established principally through their gravitational effects and its amount in the Universe has also been well determined by the PLANCK observations~\cite{planck2018}. Although dark matter is all-pervading in the Universe, their direct evidence in the laboratory is yet to be established mainly because of its no or very weak interaction with other known fundamental particles. The indirect detection of dark matter is based on the principle of detecting Standard Model particles produced due to the self-annihilation of dark matter in a suitable environment. These annihilation products can appear as the excess of the expected flux which could not be explained by other known astrophysical processes. These annihilation products could be $\gamma$-rays, neutrinos, $q\bar{q}$, or lepton anti-leptons. 
The satellite-borne experiment AMS-02 (Alpha Magnetic Spectrometer) onboard the International Space Station (ISS) that looks for anti-matter in the Universe, has reported an excess of positron-fraction beyond the positron energy 10 GeV ~\cite{Aguilar2013}. The predecessor of the AMS-02 experiment, namely PAMELA ~\cite{pamela_Nature_positron,pamela_2010statistical} also reported similar excess of positrons beyond positron energy of 10 GeV. The present AMS-02 data-set~\cite{Aguilar2014,AMS_2019_positrons,AMS_2019_electrons} measured up to 800 GeV indicates that the positron-fraction decreases with positron energy up to about 10 GeV, a phenomenon that can be explained from the behaviour of cosmic rays. But beyond 10 GeV, the data show a marked increase in the positron-fraction which appears to peak around the positron energy of 320 GeV. This increase appears to indicate the contribution of additional sources of positrons which may come from dark matter annihilation or decay, and/or from astronomical objects such as pulsars and supernova remnants~\cite{Linares_ChunCite,Evoli_Galactic20} which might be tested by further observations like HAWC~\cite{HAWC2017}. 

The dark matter interpretation of the positron excess had been pursued earlier by many authors, and after AMS-02 observations, in particular, various scenarios and models have been studied ~\cite{feng2014ams,dev2014neutrinoAMS,YBai_2018_PF,basudas2019galactic,CholisandHooperDM,IbarraSilkDM_AMS,jin2013implications,Peyman2015,Lin:2014vja,Ding_HESS_exp} either assuming an appropriate boost factor or considering the Sommerfeld effect. Models involving dark matter decay ~\cite{Siqueira_RNu,Siqueira_2ComDM,KhlopovDMDecay,Klopov2014,Farzan:2019qdm} as an explanation for the AMS-02 reported positron excess were also addressed by many authors.

In the present work, we propose a specific dark matter model by extending the Standard Model of Particle Physics by an additional scalar doublet, a singlet fermion, and also a singlet boson. We explicitly work out the phenomenology of this model to establish the singlet fermion to be our dark matter candidate. We then calculate the positron excess from the annihilation of this fermionic dark matter after incorporating the light boson mediated Sommerfeld enhancement. 
The model in fact reduces to a Two Higgs Doublet Model (2HDM) with an additional singlet boson as a portal to the fermion dark matter~\cite{Kim_fermDM08}.  The singlet boson, which is taken to be light and responsible for Sommerfeld enhancement,  mixes with the pseudoscalar inherently present in the 2HDM.  Thus,
the dark matter candidate annihilates to the singlet boson which subsequently decays to quarks and leptons.

As is well-known, the Sommerfeld enhanced annihilation to quarks and leptons are strongly constrained using the measurements of Cosmic Microwave Background Radiation (CMBR) anisotropies by PLANCK~\cite{planck2018} and the dwarf galaxy results from Fermi-LAT~\cite{FermiLAT2018}. Such a difficulty can be circumvented by making the singlet boson leptophilic and forbidding the tau final state kinematically.  
This requires the bosonic portal particle to be very light, which suits for  appropriate Sommerfeld enhancement as was suggested originally in the kinetic portal scenario~\cite{AHamed_SE,Chun_2008_U1}. Thus the dark matter annihilation is supposed to produce four muons, which is shown to provide good fits to the positron excess and spectrum data obtained by the AMS-02 experiment
for the dark matter mass around 1.6 TeV and the annihilation cross-section $\langle\sigma v\rangle\sim 2\times 10^{-23}$ cm$^3$/sec where a local boost factor $\sim 8$ has been adopted.
The leptophilic nature can be realized naturally in the lepton-specific 2HDM with large $\tan\beta$, which is known to accommodate the muon g-2 deviation~\cite{Broggio2014g-2value}.  The additional light singlet boson can 
contribute to the muon g-2 as well as other flavour observables. We also analyze, in this work, such an effect to identify
the favourable parameter space.  
Let us remark that observation of gamma rays from the galactic centre can provide another significant constraint on the dark matter annihilation~\cite{Abdallah2016_HESS,DM_HESS+Ferm+Planck}. However, 
such an observation strongly depends on the dark matter profile~\cite{Cirelli_JCAP_2011}, which have large uncertainties, and thus is not considered in this work.

The paper is organised as follows. In Section \ref{leptoframe} we describe the proposed particle dark matter model and discuss particle physics phenomenology of the model, in particular, the impact of the extra singlet boson added to the lepton-specific 2HDM which explains the observed muon g-2. Section \ref{Simple} is devoted to the discussion of dark matter relic density, Sommerfeld enhancement and the dark matter mass and couplings relevant for the calculation of the positron excess. The calculations of the positron fraction and positron spectrum, and its variations with positron energy, using the theoretical framework proposed in the work, are performed in Section \ref{sec:PositronCalc}. $\chi^2$ analyses are also performed in Section \ref{sec:PositronCalc} for the AMS-02 experimental results with the calculated results for positron fraction excess and positron spectrum. Finally, some concluding remarks are presented in Section \ref{sec:Conclu}.

\section{The Model for Leptophilic Portal}\label{leptoframe}

In this section, we describe the complete particle dark matter model proposed in the present work. We propose a particle physics model for Dark Matter (DM) by minimally extending the Standard Model (SM) of particle physics with a fermionic singlet $\chi$ and a singlet boson $\phi$ and an extra Higgs doublet. The stability of the fermion $\chi$ (the DM candidate) is ensured by imposing a $Z_2$ symmetry under which $\chi$ is odd while the SM sector is even. The singlet boson acts as a mediator between the DM and the 2HDM having the following interaction
\cite{Dorival2017,YangPRD94_DAMA,arcadi2018,dolan2015taste,bauer2017Pseudo,Bauer2017Portal2,SeydaIpek2014,JoseMiguel2016,ArcadiPortal}
\begin{equation}
\mathcal{L}_{int}=\phi \bar{\chi}(g
_{\chi}+iy_{\chi}\gamma^5)\chi + \phi\left( i b_\phi \Phi_1^\dagger \Phi_2+ {\rm h.c.}\right) 
\end{equation}   
where $b_\phi$ is a dimension-one parameter, $g_{\chi}$ and $y_{\chi}$ are the scalar and pseudoscalar like couplings between the singlet boson $\phi$ and the DM candidate $\chi$ and $\Phi_i$ ($i$=1,2) are the two Higgs doublet fields.

In the 2HDM sector, we have two charged Higgs fields ($H^{\pm}$), two CP-even scalar fields ($h,H$), one CP-odd scalar ($A_0$) and three Goldstone bosons ($G^\pm, G^0$). The Higgs doublets, $\Phi_1$ and $\Phi_2$, are written in the form~\cite{Branco2HDM,2HDMC} 
\begin{equation}
\Phi_1=
\left(
\begin{matrix}
c_\beta G^+ - s_\beta H^+\\
\frac{1}{\sqrt{2}}\left(v_1 + c_\alpha H - s_\alpha h + i c_\beta G - i s_\beta A_0\right)
\end{matrix}\right),
\end{equation}
\begin{equation}
\Phi_2=
\left(
\begin{matrix}
s_\beta G^+ + c_\beta H^+\\
\frac{1}{\sqrt{2}}\left(v_2 + s_\alpha H + c_\alpha h + i s_\beta G + i c_\beta A_0\right)
\end{matrix}\right).
\end{equation}
Here $\alpha$ is the mixing angle between the two CP even scalars and $c_x$ and $s_x$ ($x=\alpha,\beta$) represent $\cos x$ and $\sin x$ respectively.  We consider the lepton-specific 2HDM where the leptons acquire their masses from their Yukawa couplings with the $\Phi_1$ doublet while quarks get their masses due to their Yukawa couplings with the $\Phi_2$ doublet. Then the Yukawa couplings of the physical Higgs fields can be written as
\begin{equation}\label{YukawaChun}
\begin{split}
-\mathcal{L}_{Yuk} =\hspace{2mm}&\sum_{f=u,d,l} \frac{m_f}{v} \left(y^h_f h\bar{f}f + y^H_f h\bar{f}f - iy^{A_0}_f A_0\bar{f}\gamma^5f\right) \\
& + \left[\sqrt{2}V_{ud} H^+\bar{u}\left(\frac{m_u}{v} y^{A_0}_u P_L + \frac{m_d}{v}y^{A_0}_d P_R\right)d + \rm h.c. \right]\\
& + \left[\sqrt{2}\frac{m_l}{v}y^{A_0}_l H^+\bar{\nu}P_R l + \rm h.c. \right],
\end{split}
\end{equation}
where $y^{h,H,A_0}_f$ are the normalized Yukawa couplings of fermions and $v =\sqrt{v^2_1 + v^2_2}= 246\hspace{2 mm} \rm GeV$ is the total Higgs vacuum expectation value. Here the CP even scalar $h$ denotes the SM-like Higgs boson with mass $m_h=$125 GeV and $H$ is the additional Higgs boson with heavier mass $m_H$.
The normalized Yukawa couplings for only the CP-odd Higgs boson are given as
\begin{align}
y^{A_0}_u= \cot\beta,\quad    y^{A_0}_d&= -\cot\beta,\quad     y^{A_0}_l= \tan\beta
\end{align}
where $\tan\beta\equiv v_2/v_1$.

Notice that we have introduced only the pseudoscalar coupling between the singlet boson $\phi$ and $\Phi_{1,2}$. As a result, $A_0$ mixes with $\phi$ and the mass-squared squared matrix can be written as
\begin{equation}\label{MassCP}
\mathcal{M}^2_A=\left(
\begin{matrix}
m^2_{A_0} & \hspace{4mm}b_\phi v \\
b_\phi v & \hspace{4mm}m^2_\phi
\end{matrix} \right).
\end{equation}
Diagonalisation of $\mathcal{M}^2_A$ by the rotation of the state
$\left(\begin{matrix}
A_0\\
\phi
\end{matrix}\right)$ as
\begin{equation}\label{PseudoMix}
\left(
\begin{matrix}
A\\
a
\end{matrix}\right)=
\left(
\begin{matrix}
\cos\theta & \sin\theta\\
-\sin\theta & \cos\theta
\end{matrix}\right)
\left(
\begin{matrix}
A_0\\
\phi
\end{matrix}\right)
\end{equation}
leads to the relation
\begin{equation}
b_\phi=\frac{1}{2v}\left(m^2_A-m^2_a\right)\sin 2\theta.
\end{equation}
where $m_{A,a}^2$ are the eigenvalues of $\mathcal{M}^2_A$ for the rotated eigenstate
$\left(\begin{matrix}
A\\
a
\end{matrix}\right)$\,\,.

In terms of the physical CP-odd eigenstates $A$ and $a$, the interaction Lagrangian $\mathcal{L}_{int}$ now takes the form
\begin{equation}\label{modInteraction}
\mathcal{L}_{int}\supset(A\sin\theta + a\cos\theta)\bar{\chi}(g_{\chi}+iy_{\chi}\gamma^{5})\chi\,\,.
\end{equation}
Accordingly, the Yukawa Lagrangian in Eq. (\ref{YukawaChun}) modifies as
\begin{equation}\label{modYukwa}
\begin{split}
-\mathcal{L}_{Yuk}\supset-i\sum_{f=u,d,l} \frac{m_f}{v} y^{A_0}_f (\cos\theta A - \sin\theta a)\bar{f}\gamma^5f .
\end{split}
\end{equation}
From Eqs. (\ref{modInteraction}-\ref{modYukwa}), the couplings of the physical states $a$ and $A$ to quarks, leptons and DM fermion can be wriiten as 
\begin{align}
&y^a_l=-\tan\beta\sin\theta, && y^A_l=\tan\beta\cos\theta,
\nonumber \\
&y^a_u=-\cot\beta\sin\theta, &&y^A_u=\cot\beta\cos\theta,  \nonumber\\
&y^a_d=\cot\beta\sin\theta,  && y^A_d=-\cot\beta\cos\theta, \\
&g_{a\bar{\chi}\chi}=g_\chi\cos\theta,  &&g_{A\bar{\chi}\chi}=g_\chi\sin\theta, \nonumber\\
&y_{a\bar{\chi}\chi}=y_\chi\cos\theta, &  &y_{A\bar{\chi}\chi}=y_\chi\sin\theta. \nonumber
\end{align}
Having assumed CP invariance in the 2HDM sector,  the physical CP-odd state $a$ (and also $A$) 
does not have any couplings of the type $aZZ$, $aW^+W^-$ or $ahh$.

\subsection{LHC constraints}

The lepton-specific 2HDM  becomes leptophilic at large $\tan\beta$ (called Leptophilic-2HDM or L2HDM) and thus easily avoids constraints from hadronic processes. The extra Higgs bosons $a, A, H$ and $H^\pm$ in the present model can hardly be produced only through the quark couplings at the LHC. However, one can still probe their existence through the Drell-Yann process~\cite{Chun_HRI2018} or the Standard Model Higgs decays, $h\to aa, aA, AA$~\cite{Chun_HRI_PLB2017} if allowed kinematically.  
In the present work, the pseudoscalars $a$ and $A$ are required to be very light $m_a \ll m_A < m_h/2$ and thus their production through $h\to \phi \phi'$ ($\phi,\phi'= a, A$) may be observed if the corresponding coupling $g_{h\phi\phi'}$ (see~\cite{abe2019loop} for the full expression) is large enough.
 The decay width for such a process is given by
\begin{equation}\label{Decayhtoaa}
\Gamma_{h\rightarrow \phi\phi'}=S\frac{g_{h\phi\phi'}^2}{32\pi m_h} \sqrt{1-\frac{(m_{\phi}+m_{\phi'})^2}{m_h^2}}\,\,,
\end{equation}
where $S=2$ for $\phi \neq \phi'$. Indeed such channels are examined by the LHC experiments.
Considering the $2\mu2\tau$ and $4\mu$ searches  at CMS~\cite{2mu2tau2018search,4mu}, we obtain the following limits
\begin{eqnarray}
g_{hAA}/v &\lesssim& 0.008 ~\mbox{for}~  m_A= 15-62\,\, \mbox{GeV}, \\
g_{haa} /v &\lesssim& 0.003~\mbox{for}~ m_a=0.5-3.5\,\, \mbox{GeV},
\end{eqnarray}
applicable to the mass ranges under consideration.

\subsection{Flavour constraints}
\begin{figure}[h]
	\centering
	\includegraphics[scale=0.65]{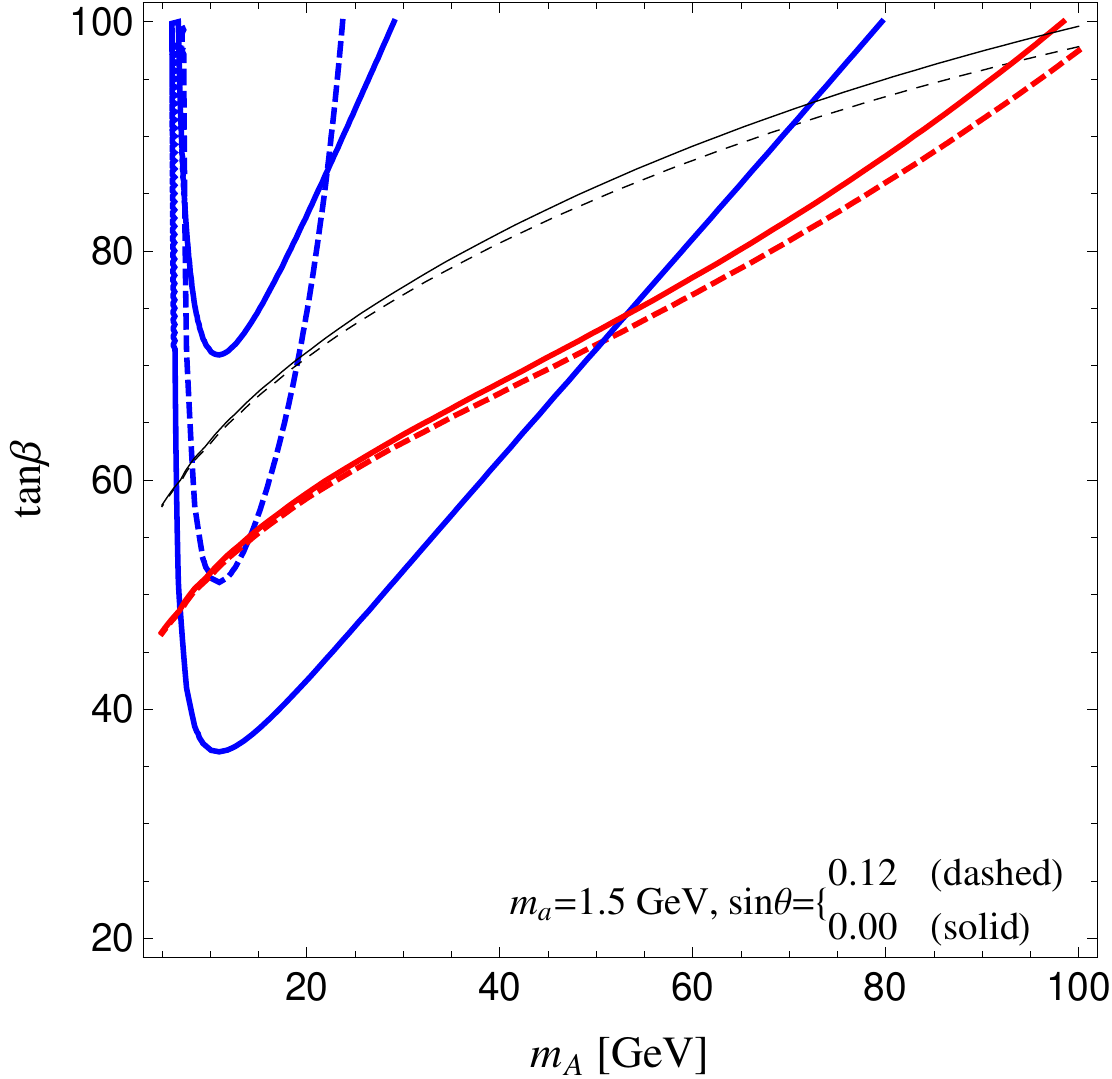}
	\includegraphics[scale=0.65]{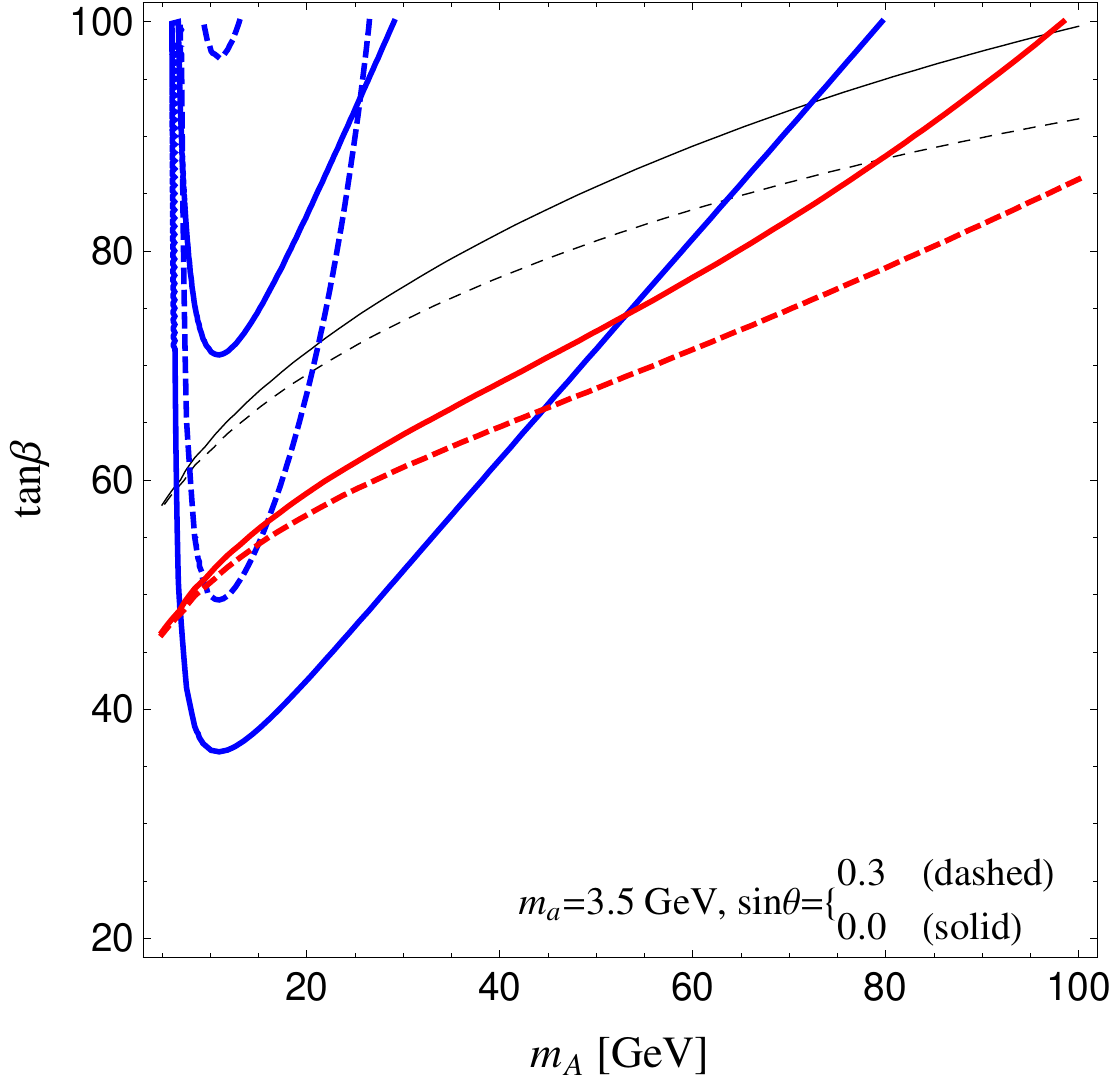}
	\caption{The combined 2$\sigma$ allowed parameter space in the $\tan\beta$ vs $m_A$ plane with $m_a=1.5$ GeV (left panel) and $m_a=3.5$ GeV (right panel) from the $a_{\mu}$ results (blue) and upper limits from the lepton-flavour universality bounds in the $Z$ (red) and $\tau/\mu$ (black) decays. We have assumed $m_H=m_{H^{\pm}}=300$ GeV.  } 
	\label{muong2}
\end{figure}

As is well-known, the L2HDM can explain the muon g-2 anomaly due to the presence of a light pseudo scalar $A$~\cite{Broggio2014g-2value,Cao2009g-2} whose parameter space is constrained by the $B_s \to \mu^+ \mu^-$ decay ~\cite{Wang2014g-2} and lepton universality conditions in $Z\to ll$ and $\tau/\mu$ decays~\cite{abe2015lepton}. The presence of  another (lighter) pseudoscalar $a$ can modify the previous results and 
puts a limit in its parameter space. Generalizing the calculations in \cite{chun2016leptonic}, we update the 
conventional parameter region of ($m_A, \tan\beta$) favoured by the recent measurement of $a_\mu=(g-2)_\mu/2$ at 
FNAL~\cite{Muong-2_FNAL}:
$$a_{\mu}^{\rm Exp.}-a_\mu^{\rm Th.} =251(59)\times 10^{-11}$$ 
and examine the impact of $a$.
In Fig.~\ref{muong2},
we show the region of parameter space consistent with
the 2$\sigma$ range of the muon g-2 data when $m_{H^\pm}=m_H=300$ GeV is adopted. The region inside the solid blue contours is for the pure L2HDM. In this case there is negligible mixing with $a$ signifying $\sin\theta\to0$. When the mixing is turned on, the light pseudoscalar gives a negative contribution to the muon g-2 and thus large $\tan\beta$ and smaller $m_A$ are required to explain the muon g-2 as shown by the dashed blue lines in both the left and right panels. This leads to the upper limit on the mixing angle
\begin{equation}
 \sin\theta=0.12-0.3\,\,,
\end{equation}
for the preferred range of $m_a$ ($m_a=1.5-3.5$ GeV) as will be discussed in the following sections. A light pseudoscalar can serve well as a mediator of $B_s$ decays to a pair of leptons. Its impact especially on the $B_s\rightarrow\mu^+\mu^-$ decay channel can be sizable due to its ``tree-level'' exchange accompanied by loop-level flavour changing neutral currents. Recently, the LHCb collaboration~\cite{LHCbBs} reported the  $B_s\rightarrow\mu^+\mu^-$ branching ratio to be ${\rm Br}(B_s\rightarrow\mu^+\mu^-)=(3.0\pm 0.6_{-0.2}^{+0.3})\times 10^{-9}$ which is largely consistent with the Standard Model prediction. We do not attempt to revise the analysis accordingly but quote the previous results
constraining the mediator mass $m_A \gtrsim 10$ GeV~\cite{Wang2014g-2} which can be translated to $m_a \gtrsim 10 \sin\theta$ GeV. 

\subsection{Searches for a light leptophilic boson}\label{SecSearch}
\begin{figure}[h]
	\centering
	\includegraphics[scale=0.95]{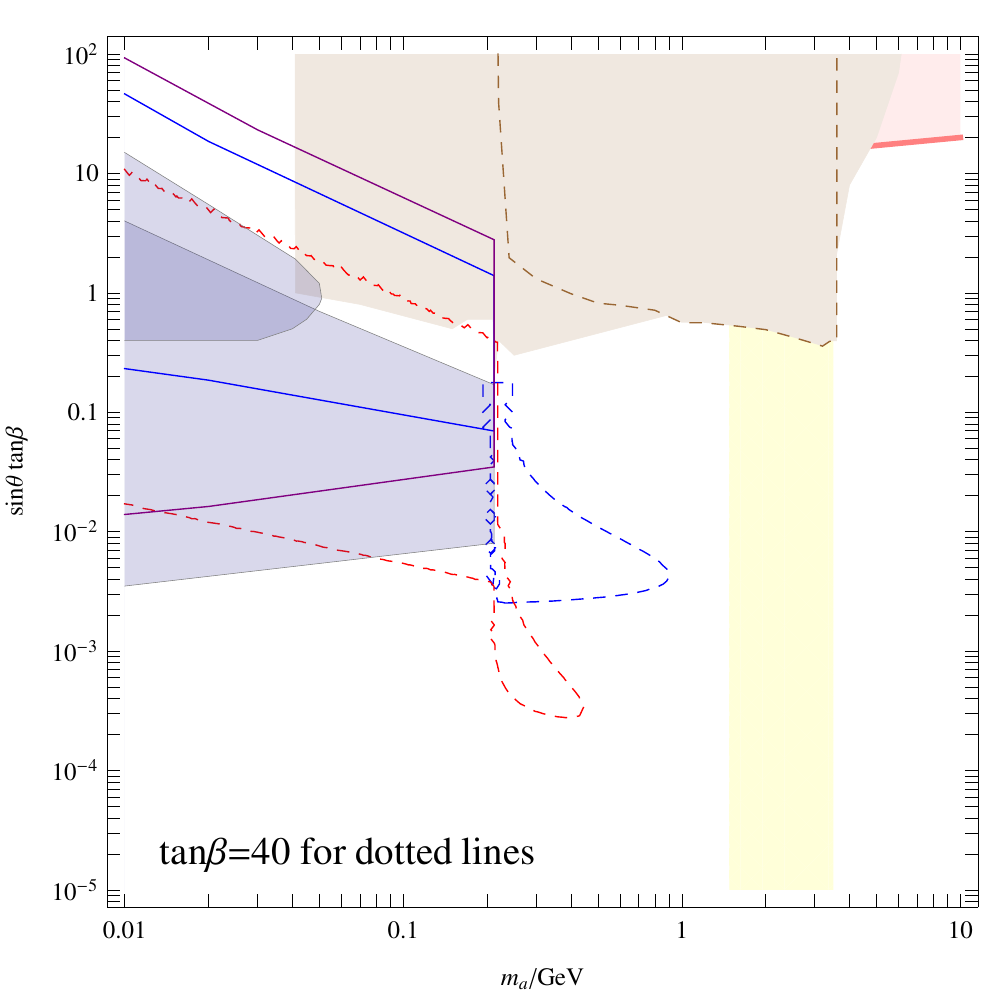}
	\caption{The shaded regions are excluded by BABAR (light brown), ORSAY and E137 (light blue), and the solid contours are future sensitivities at NA64-muon (purple), SEAQUEST (blue) and lepton colliders (pink on the right upper corner). The shaded region inside the dotted line is excluded by LHCb (light brown), and the other dotted contours are future sensitivity at BELLE II (blue) and SHIP/MATHUSLA (red). 
	}
	\label{FigSearch}
\end{figure}

There have been attempts to search for new light particles that have been predicted to couple to Standard Model in various theoretical frameworks~\cite{Flavor1}. Extending previous studies on leptophilic bosons~\cite{Flavor2,Flavor3}, we further obtain the current constraints and future sensitivity on the parameter space of our leptophilic pseudoscalar. Its lepton coupling proportional to $\sim\theta \tan\beta$ can be probed directly by lepton processes at ORSAY~\cite{Orsay}, E137~\cite{E137_Bjorken}, and BABAR~\cite{BaBar_Results}, as well as the future experiments such as NA64-muon~\cite{na64}, SEAQUEST~\cite{SeaQuest}, and lepton colliders~\cite{Chun_LC2019,Chun_LC2021}. 
The current and future search limits are presented by solid lines in Fig.~\ref{FigSearch}.
The three shaded areas are excluded by ORSAY, E137 and BABAR data, respectively. 
As the quark coupling is suppressed by 1/$\tan^2\beta$ compared to the lepton coupling, hadronic processes are inefficient to probe such a leptophilic boson. We show the current and future limits by three dotted contours for LHCb ~\cite{LHCb_2}, BELLE II~\cite{Belle2}, and SHIP~\cite{Ship} (overlapping almost with MATHUSLA~\cite{Mathusla}). For this we considered  the dominant process of $b\to s\, a (\mu\mu)$ fixing $\tan\beta=40$.  The Yellow band shows our preferred region of $m_a\approx 1.5-3.5$ GeV.

\section{Dark Matter Phenomenology}\label{Simple}

The interaction Lagrangian for the dark matter candidate and its portal in the framework of the present model (see Section \ref{leptoframe}), relevant for our analysis, can be written as 
\begin{equation}\label{Lag_simp}
\mathcal{L}=
a\bar{\chi}(g_{a\chi\chi}+iy_{a\chi\chi}\gamma^5)\chi+i a\sum_f \frac{m_l}{v} y_l^a \bar{l}\gamma^5 l
\end{equation}
where $g_{a\chi\chi}=g_\chi \cos\theta$, $y_{a\chi\chi}=y_\chi \cos\theta$, and $y_l^a=\tan\beta \sin\theta$ for $l=e,\mu,\tau$.  Note that we have omitted the quark coupling $y_q^a=\sin\theta/\tan\beta=y_l^a/\tan^2\beta$
which is highly suppressed for large $\tan\beta$ under consideration.
To study the dark matter properties, we will restrict ourselves to the small mixing limit $\sin\theta \approx 0$ for which we can take the approximation $g_{a\chi\chi}\approx g_\chi$ and $y_{a\chi\chi}\approx y_\chi$.

\subsection{Dark Matter Annihilation and Relic Density} \label{SecRel}

\begin{figure}
    \centering
    \begin{tikzpicture}[baseline={(current bounding box.center)}]
    \begin{feynman}
    \vertex (a) {\(\chi\)};
    \vertex[right=1.8cm of a] (b);
    \vertex[right=1.5cm of b] (c) {\(a\)};
  
    \vertex[below=2cm of a] (d) {\(\bar{\chi}\)};
    \vertex[right=1.8cm of d] (e);
    \vertex[right=1.5cm of e] (f) {\(a\)};
    \diagram* {
        (a) -- [fermion] (b) -- [scalar] (c),
        (b) -- [fermion] (e),
        (f) -- [scalar] (e) -- [fermion] (d),
    };
    \vertex [above=0.5em of b] {\(g_{\chi}\)};
    \vertex [below=0.5em of e] {\(y_{\chi}\)};  
    \end{feynman}
    \end{tikzpicture}
    \hspace{5mm}
    \begin{tikzpicture}[baseline={(current bounding box.center)}]
    \begin{feynman}
    \vertex (a) {\(\chi\)};
    \vertex[right=1.8cm of a] (b);
    \vertex[right=1.5cm of b] (c) {\(a\)};
  
    \vertex[below=2cm of a] (d) {\(\bar{\chi}\)};
    \vertex[right=1.8cm of d] (e);
    \vertex[right=1.5cm of e] (f) {\(a\)};
    \diagram* {
        (a) -- [fermion] (b) -- [scalar] (c),
        (b) -- [fermion] (e),
        (f) -- [scalar] (e) -- [fermion] (d),
    };
    \vertex [above=0.5em of b] {\(g_{\chi}\)};
    \vertex [below=0.5em of e] {\(g_{\chi}\)};  
    \end{feynman}
    \end{tikzpicture}
    \hspace{5mm}
    \begin{tikzpicture}[baseline={(current bounding box.center)}]
    \begin{feynman}
    \vertex (a) {\(\chi\)};
    \vertex[right=1.8cm of a] (b);
    \vertex[right=1.5cm of b] (c) {\(a\)};
  
    \vertex[below=2cm of a] (d) {\(\bar{\chi}\)};
    \vertex[right=1.8cm of d] (e);
    \vertex[right=1.5cm of e] (f) {\(a\)};
    \diagram* {
        (a) -- [fermion] (b) -- [scalar] (c),
        (b) -- [fermion] (e),
        (f) -- [scalar] (e) -- [fermion] (d),
    };
    \vertex [above=0.5em of b] {\(y_{\chi}\)};
    \vertex [below=0.5em of e] {\(y_{\chi}\)};  
    \end{feynman}
    \end{tikzpicture}
    \caption{The diagrams for DM annihilation into $aa$.}
    \label{DMAnnihilation}
\end{figure}
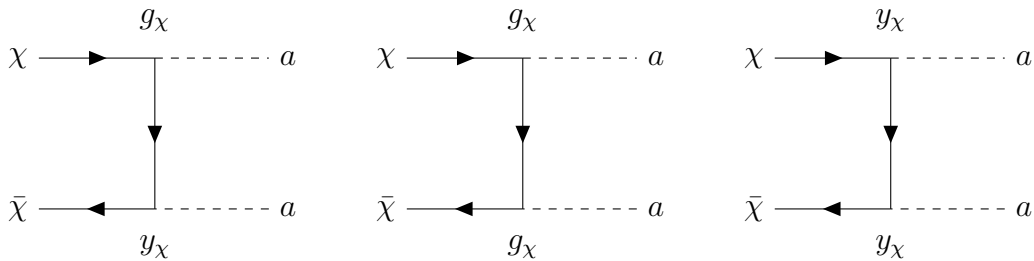

The Feynman diagrams for the main dark matter annihilation channels in this framework are shown in Fig.~ \ref{DMAnnihilation}. The total annihilation cross-section is given as,
\begin{equation}\label{tchanann}
\langle\sigma v_{rel}\rangle_{\bar{\chi}\chi\rightarrow aa}\simeq\left[ \frac{g^2_{\chi}y^2_{\chi}}{8\pi m^2_{\chi}}+\frac{3 g^4_{\chi}}{64\pi m^2_{\chi}}\langle v^2\rangle+\frac{y^4_{\chi}}{384\pi m^2_{\chi}}\langle v^2\rangle\right]\left(1-\frac{m^{2}_{a}}{m^{2}_{\chi}}\right)^{1/2}\,\,.
\end{equation}
It can be seen that the annihilation due to the second and third t-channel processes suffer a p-wave suppression of order $\mathcal{O}(v^2_{rel})$~\cite{BellPRD.96_Tchan,2HDM+Scalar_Dutta}. Although at freeze-out temperature of $\chi$ ($v_{rel}\sim0.3$) the contribution to the dark matter relic density from these processes cannot be neglected, they are highly suppressed at the present epoch ($v_{rel} \sim 10^{-3}$). 
As a result, observable signals will appear only through the first process which will be responsible for
the positron excess. 
\begin{figure}[h]
	\centering
	\includegraphics[scale=0.4]{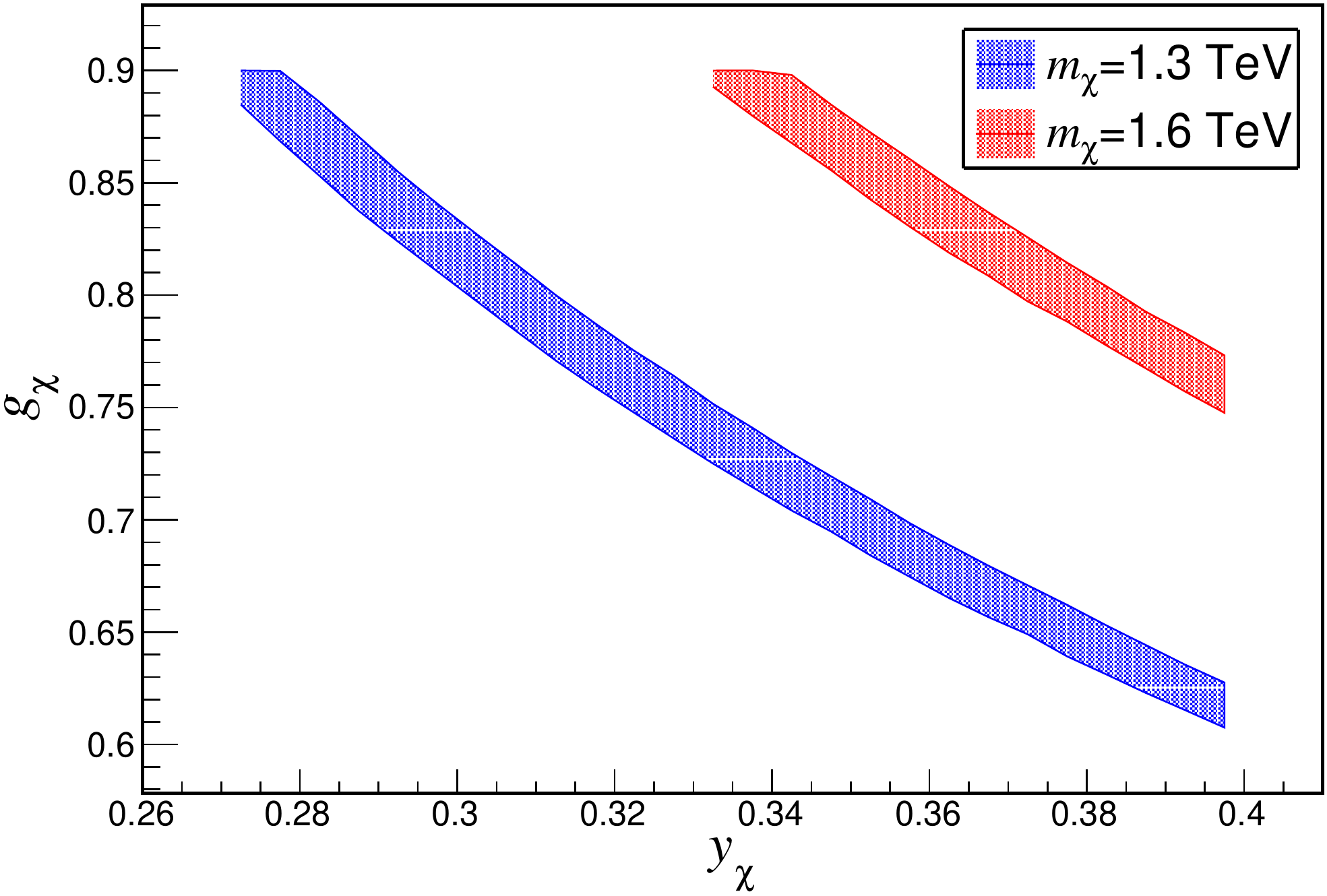}
	\caption{The allowed parameter region in the $g_{\chi}$ vs $y_{\chi}$ plane that satisfies the PLANCK observed DM relic density for dark matter masses of 1.3 TeV and 1.6 TeV. 
	}
	\label{DM_Coup}
\end{figure}

Following the standard calculation of the DM freeze-out process~\cite{kolb1990early}, one can constrain the parameters $m_\chi$, $g_\chi$ and $y_\chi$ leading to the observed DM relic density which requires $\langle\sigma v_{rel}\rangle \approx 3\times 10^{-26}$ cm$^3$/sec. In Fig. \ref{DM_Coup} we show the variation of the scalar $g_{\chi}$ as a function of pseudoscalar like coupling $y_{\chi}$, for dark matter masses 1.3 TeV and 1.6 TeV respectively, in order to agree with the dark matter relic density obtained from the PLANCK experimental results. Three sets of parameters designated by Set I. Set II and Set III are chosen from the allowed parameter space with $y_{\chi}=0.4$. These sets are given in Table \ref{TestPoints}. These will be used to obtain appropriate Sommerfeld enhancement factors and then performing the $\chi^2$ fit to the positron excess in the following sections.
\begin{table}
	\begin{center}
		  \scalebox{1}{
				\begin{tabular}{|m {1.5 cm}|m {2 cm}|m {2 cm}|m {1.5 cm}|m {1.5 cm}|}
					\hline
					& $m_\chi$(TeV)& $m_a$(GeV) & $g_\chi$ & $y_{\chi}$ \\
					\hline
					Set I & 1.3 & 1.97  & 0.60 & 0.4  \\
					\hline
					Set II & 1.6 & 3.42  & 0.75& 0.4  \\
					\hline
					Set III & 1.9 & 3.45   & 0.88 & 0.4  \\
					\hline
			\end{tabular}}
		\caption{The representative sets of parameters leading to the observed DM relic density and appropriate Sommerfeld enhancement.}
			\label{TestPoints}   
	\end{center}
\end{table}

\subsection{Direct Detection of Dark Matter}

\begin{figure}
    \centering
    \begin{tikzpicture}[baseline={(current bounding box.center)}]
    \begin{feynman}
    \vertex (a) {\(\chi\)};
    \vertex[right=1.8cm of a] (b);
    \vertex[right=1.5cm of b] (c) {\(\chi\)};
  
    \vertex[below=2cm of a] (d) {\(q\)};
    \vertex[right=1.8cm of d] (e);
    \vertex[right=1.5cm of e] (f) {\(q\)};
    \diagram* {
        (a) -- [fermion] (b) -- [fermion] (c),
        (b) -- [scalar, edge label'=\(a\)] (e),
        (d) -- [fermion] (e) -- [fermion] (f),
    };
    \vertex [above=0.5em of b] {\(g_{a\chi\chi}, i y_{a\chi\chi} \gamma_5\)};
    \vertex [below=0.5em of e] {\(ig_{aqq}\gamma_5\)};  
    \end{feynman}
    \end{tikzpicture}
    \hspace{5mm}
    \begin{tikzpicture}[baseline={(current bounding box.center)}]
    \begin{feynman}
    \vertex (a) {\(\chi\)};
    \vertex[right=1.5cm of a] (b);
    \vertex[right=1.0cm of b] (c);
    \vertex[right=1.2cm of c] (g) {\(\chi\)};
  
    \vertex[below=2cm of a] (d) {\(q\)};
    \vertex[right=1.5cm of d] (e);
    \vertex[right=1.0cm of e] (f);
    \vertex[right=1.2cm of f] (h) {\(q\)};
    \diagram* {
        (a) -- [fermion] (b) -- [fermion] (c) -- [fermion] (g),
        (b) -- [scalar, edge label'=\(a\)] (e), (c) -- [scalar, edge label=\(a\)] (f),
        (d) -- [fermion] (e) -- [fermion] (f) -- [fermion] (h),
    };
    \end{feynman}
    \end{tikzpicture}
    \hspace{5mm}
    \begin{tikzpicture}[baseline={(current bounding box.center)}]
    \begin{feynman}
    \vertex (a) {\(\chi\)};
    \vertex[right=1.5cm of a] (b);
    \vertex[right=1.0cm of b] (c);
    \vertex[right=1.2cm of c] (g) {\(\chi\)};
  
    \vertex[below=2cm of a] (d) {\(q\)};
    \vertex[right=2.0cm of d] (e);
    \vertex[above=1.3cm of e] (h);
    \vertex[right=1.8cm of e] (f) {\(q\)};
    \diagram* {
        (a) -- [fermion] (b) -- [fermion] (c) -- [fermion] (g),
        (b) -- [scalar, edge label'=\(a\)] (h), (c) -- [scalar, edge label=\(a\)] (h), (h) -- [scalar, edge label'=\(h\,/H\)] (e),
        (d) -- [fermion] (e) -- [fermion] (f),
    };
    \end{feynman}
    \end{tikzpicture}
    \caption{The dominant tree and loop (box and triangle) diagrams for DM direct detection. The coupling are given by $g_{a\chi\chi}=g_\chi \cos\theta$,  $y_{a\chi\chi}=y_\chi \cos\theta$, and  $g_{aqq} = (m_q/v) \sin\theta/\tan\beta$}
    \label{SimDD}
\end{figure}
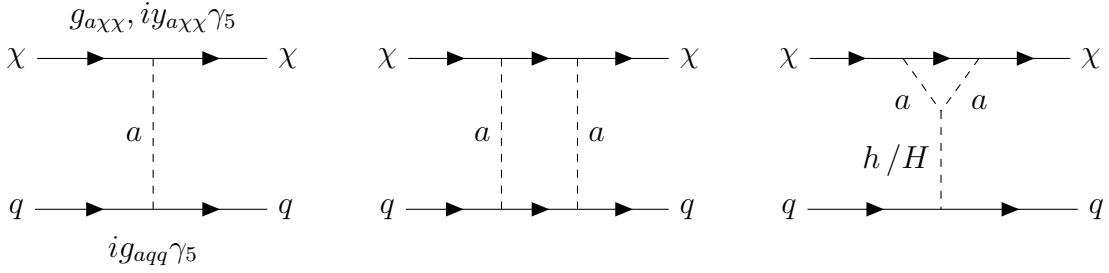

In the present work, with a leptophilic pseudoscalar mediator to quarks, the tree-level process for the DM-nucleon scattering is spin-dependent and its cross-section, being proportional to $\vec{q}^2/m^2_{\chi}$ ($\vec{q}$ is a momentum-transfer) and the quantity $\sin^2\theta/\tan^2\beta$, is too much suppressed to be observable in the foreseeable future.  There are also one-loop processes through the box and triangle diagrams \ref{SimDD} which give rise to spin-independent DM-nucleon scattering  \cite{arcadi2018,abe2019loop}. In our case, however, the box diagram is proportional to $\sin^2\theta/\tan^2\beta$ and the triangle diagram involves the small couplings $g_{haa}/v \lesssim 0.003$ and  $y_{Hqq}\approx \frac{m_q}{v\,\tan\beta}$ in amplitudes. As a result, we find the corresponding cross-sections are below the neutrino forward scattering cross-section limit for $m_\chi \sim 1$ TeV. Therefore, the dark matter candidate in the present model may not be probed by direct detection experiments. 

\subsection{Sommerfeld Enhancement}\label{sec:SECalc}

In the early Universe, at around the freeze-out temperature of the thermal relics, the relative velocity of DM particles was relativistic with $v_{rel}\sim0.3$.  The annihilation rate and the relic abundance were thus determined by this relative velocity. As discussed above the required annihilation rate for DM particles to  reproduce the observed relic abundance is $\left\langle\sigma v_{rel}\right\rangle\simeq3.0\times10^{-26}$ $\rm cm^3/sec$.  In the present Universe, however, the DM particles are highly non-relativistic with $v_{rel}\sim10^{-3}$, and thus the DM annihilation can get enhanced considerably by the Sommerfeld effect~\cite{sommerfeld1931}.
The thermally averaged annihilation cross-section in the non-relativistic limit can be written as
\begin{equation}\label{ThermAnnNR}
\left\langle \sigma v_{rel}\right\rangle=\sqrt{\frac{2}{\pi}}\frac{1}{v^3_0}\int v^2_{rel}e^{-(v^2_{rel}/2v^2_0)}\sigma v_{rel}dv_{rel}\,\,,
\end{equation}
where $v_0\simeq 220\hspace{1mm} \rm km/sec$ is the most probable velocity of dark matter in the galaxy. One can also write the most probable velocity as
\begin{equation}
v_0=\sqrt{\frac{2}{x_{\chi}}}\,\,,
\end{equation}
where $x_{\chi}=m_{\chi}/T_{\chi}$, $T_{\chi}$ being the temperature of dark matter today. We can also write the Sommerfeld enhanced annihilation rate $\sigma v_{rel}$ as $ (\sigma v_{rel})_0S_0(v_{rel})$ where $S_0(v_{rel})$ is the velocity-dependent Sommerfeld enhancement factor.  Then, one obtains ~\cite{L.Feng_SE}
\begin{equation}\label{Savg}
\begin{split}
\left\langle \sigma v_{rel}\right\rangle&=\frac{x^{3/2}_{\chi}}{2\sqrt{\pi}}\int_{0}^{v_{esc}}(\sigma v_{rel})v^2_{rel}e^{-(x_{\chi}v^2_{rel}/4)}dv_{rel}\\
&=(\sigma v_{rel})_0\left\langle S\right\rangle,
\end{split}\,\,,
\end{equation}
where $v_{esc}\simeq 550\hspace{1mm}\rm km/sec$ is the escape velocity for dark matter in the DM halo of our galaxy and 
\begin{equation}\label{Seff_therm}
\left\langle S\right\rangle=\frac{x^{3/2}_{\chi}}{2\sqrt{\pi}}\int_{0}^{v_{esc}}S_0(v_{rel})v^2_{rel}e^{-(x_{\chi}v^2_{rel}/4)}dv_{rel}
\end{equation}
stands for the thermally averaged Sommerfeld enhancement.

In the present framework, the dark matter interacts with the light CP-odd particle $a$ through both the scalar and pseudoscalar couplings. In section \ref{SecRel}, we have chosen the scalar coupling, $g_{\chi}$ to be greater than the pseudoscalar coupling $y_{\chi}$ so that the $a$-$\chi$ scalar interaction is more effective and relevant for the Sommerfeld enhancement~\cite{Agrawal:2020lea}.
In order to calculate the Sommerfeld factor, we rely on the analytic formula given by
\cite{Hulthen_Cassel_SE,Hulthen_Slatyer_SE}
\begin{equation}\label{S0_analytical}
S_0=\frac{\pi}{\epsilon_v}\hspace{1mm}\frac{\sinh\left(\frac{2\pi\epsilon_v}{\pi^2\epsilon_a/6}\right)}{\cosh\left(\frac{2\pi\epsilon_v}{\pi^2\epsilon_a/6}\right)-\cos\left(2\pi\sqrt{\frac{1}{\pi^2\epsilon_a/6}-\frac{\epsilon^2_v}{\left(\pi^2\epsilon_a/6\right)^2}}\right)}\,\,,
\end{equation}
where $\epsilon_v=v_{rel}/\alpha_{\chi}$ and $\epsilon_a=m_a/\alpha_{\chi}m_{\chi}$. This analytical solution is known to match well, within 10\% variation, with the numerical solutions~\cite{L.Feng_SE}.
It is now crucial to require $\epsilon_v < \epsilon_\alpha$ saturating the velocity depenence and obtain the Sommerfeld-enchanced annihilation rate around $10^{-24}$ cm$^3$/sec (that is, $\langle S \rangle \sim 100$) allowed by the CMB limit for the annihilation process $\chi\chi\to a a \to 4 \mu$. Thus, we are forced to be in a narrow range of the light boson mass, 
 $1.5\hspace{1mm}{\rm GeV} < m_a < 3.5\hspace{1mm}{\rm GeV}$, as shown in Table \ref{TestPoints}.
We will see that the dark matter mass needs to be in the range,  $1.3$ TeV$\le m_{\chi}\le1.9$ TeV, to fit the positron excess with an additional boost factor of $B\sim 8$.

\section{Positron Excesses}\label{sec:PositronCalc}

Having set all the relevant ingredients for the dark matter annihilation, 
we calculate the positron flux from DM annihilation and finally the positron fraction resulting from 
the dominant channel   $\bar{\chi}\chi\rightarrow a a \to 4 \mu$.
The thermally averaged cross-section of this channel in our galaxy is given by 
\begin{equation}\label{DomChCross}
	\begin{split}
	\left\langle \sigma v_{rel}\right\rangle_{\bar{\chi}\chi\rightarrow aa} \approx  \frac{g^2_{\chi}y^2_{\chi}}{8\pi m^2_{\chi}} \langle S\rangle .
	\end{split}
\end{equation}
In this work, we have chosen $m_a<2m_{\tau}$ and as a result the $a\rightarrow\tau^+\tau^-$ is kinematically forbidden. Therefore $a$ will readily decay to $2\mu$, which can finally produce positrons. 
The positron fraction as a result of DM annihilations, per unit energy, is given by ~\cite{Cirelli_Bkg}
\begin{equation}
\frac{d\Phi_{e^+}}{dE}(t,\vec{x},E)=v_{e^+}\frac{f}{4\pi}\hspace{2 mm} \rm GeV^{-1} cm^{-2} s^{-1} sr^{-1}
\end{equation}
where $v_{e^+}$ is the velocity of positron and the quantity $f=\frac{dN_{e^+}}{dE}$ represents the positron number density per unit energy. The latter follows the diffusion equation ~\cite{baltz1998positron,delahaye2009galactic}
\begin{equation}\label{Diffusion}
\frac{\partial f}{\partial t}-\nabla(\mathcal{K}(E,\vec{x})\nabla f)-\frac{\partial}{\partial E}(b(E,\vec{x})f)=Q(E,\vec{x})\,\,,
\end{equation}
where $\mathcal{K}(E,\vec{x})$ and $b(E,\vec{x})$ are the diffusion coefficient function and energy loss coefficient function respectively. The source term $Q$ in Eq. (\ref{Diffusion}) is given by ~\cite{Cirelli_JCAP_2011}
\begin{equation}
Q=\frac{1}{2}\left(\frac{\rho_{DM}}{m_{DM}}\right)^2\hspace{1 mm}f^{ann}_{inj}\,\,,
\end{equation}
where the quantity $f^{ann}_{inj}=\sum_{k}\left\langle\sigma v\right\rangle_k \frac{dN^k_{e^+}}{dE}$. Here $k$ represents all annihilation channels with $e^+$ in the final state, $\rho_{DM}$ and $m_{DM}$ represent the dark matter density and mass respectively and $\frac{dN_{e^+}}{dE}$ is the positron spectrum. The differential positron flux at Earth with positron energy $E$ from DM annihilation can be written as ~\cite{Cirelli_Bkg,Cirelli_JCAP_2011},
\begin{equation}
\frac{d\Phi_{e^+}}{dE}(E,\vec{r}_{\odot})=B\frac{v_{e^+}}{4\pi b(E,\vec{x})}\frac{1}{2}\left(\frac{\rho_{\odot}}{m_{DM}}\right)^2\sum_{k}\left\langle\sigma v\right\rangle_k \int_{E}^{m_{DM}} dE_s\frac{dN^k_{e^+}}{dE}(E_s)I(E,E_s,\vec{r}_{\odot}),
\end{equation}
where $E_s$ stands for the positron energy at production (source) and $I(E,E_s,\vec{r}_{\odot})$ is the generalized halo function at Earth, which is in fact the Green function from a source with energy $E_s$. In the above, the local dark matter density is considered to be $\rho_{\odot}\sim0.3$ GeV/$\rm cm^3$ and $B$ represents the cosmological boost factor which accounts for dark matter clumping. The positron flux at Earth, from the annihilation of the DM candidate proposed in this work, is calculated by using the publicly available code
{\fontfamily{ccr}\selectfont PPPC4DMID} ~\cite{Cirelli_JCAP_2011,BoudaudCosmicPPPC}. The boost factor $B$ depends on the energy $E$ of the positron ~\cite{lavalle2008Nbody,brun2007antiprotonpositron,berezinsky2003smlumps} with a value that lies between 1-20 ~\cite{lavalle2007clumpiness}, depending on the energy. A reasonable value of $B\sim10$ ~\cite{Cirelli_Bkg,goh2009leptonic}.
The positron fraction as a result of DM annihilations in this work is given by ~\cite{YBai_2018_PF}
\begin{equation}
F_{e^+}=\frac{\Phi^{sig}_{e^+}+\Phi^{bkg}_{e^+}}{2\Phi^{sig}_{e^+}+\Phi^{bkg}_{e^+}+\Phi^{bkg}_{e^-}}\,\,,
\end{equation}
where $\Phi^{sig}_{e^{+}}$ and $\Phi^{bkg}_{e^{\pm}}$ denote the flux of the positrons from DM annihilation and background cosmic ray flux respectively and it has been assumed that the positron and electron fluxes produced as a result of DM annihilation are same. The positron background has been computed by adopting a parametrisation of the background as ~\cite{AMS_2019_positrons}
\begin{equation}\label{AMSposBkg}
\Phi^{bkg}_{e^+}(E)=\frac{E^2}{\hat{E}^2}C_d(\hat{E}/E_1)^{\gamma_d}\,\,,
\end{equation}
where $\hat{E}=E+\phi_{e^+}$, with $\phi_{e^+}$ accounting for the solar effects. In Eq. (\ref{AMSposBkg}) above, the parameters $E_1=7.0$ GeV, $\gamma_d=-3.6$, $C_d=6.42\times10^{-2}$ $\rm GeV^{-1}m^{-2}s^{-1}sr^{-1}$ and $\phi_{e^+}=0.869$ GeV ~\cite{basudas2019galactic}, and
\begin{equation}\label{AMSelecBkg}
\Phi^{bkg}_{e^-}(E)=C_e(E/E_{1e})^{\gamma_e}\,\,,
\end{equation} 
gives the background parametrisation for electrons~\cite{AMS_2019_electrons} where $E_{1e}=42.01$ GeV, $\gamma_e=-3.3$ and $C_e=2.1\times 10^{-3}$ $\rm GeV^{-1}m^{-2}s^{-1}sr^{-1}$.
\begin{table}
	\begin{center}
	   \scalebox{1}{
			\begin{tabular}{|m {1.2 cm}|m {1.8 cm}|m {1,2 cm}|m {3 cm}|m {1.5cm}|m {3.4 cm}|}
				\hline
				& $m_{\chi}$ (TeV)  & $\langle S \rangle$ & $\left\langle \sigma v_{rel}\right\rangle \hspace{1 mm}\rm (cm^3/sec)$ &  $B$  & $B \left\langle \sigma v_{rel}\right\rangle\; \rm (cm^3/sec)$ \\
				\hline
				BP1 & 1.3  & 80 & $1.30\times10^{-24}$ & 7.02 & $0.91 \times 10^{-23}$ \\
				\hline
			    BP2 & 1.6  & 98 & $1.59\times10^{-24}$ & 8.24 & $1.31 \times 10^{-23}$ \\
				\hline
				BP3 & 1.9  & 148 & $2.39\times10^{-24}$ & 7.38 & $1.77 \times 10^{-23}$ \\
				\hline
		\end{tabular}}
		\caption{The  values of the DM mass $m_{\chi}$,  Sommerfeld enhancement $\langle S\rangle$,  the boost factor $B$, and  the annihilation cross-sections including the Sommerfeld effect and then boosted by the local factor $B$.  The values of the other parameters used in the calculation are given in Table. \ref{TestPoints}.}
		\label{SE_BP}   
	\end{center}
\end{table}
\begin{figure}[h]
	\centering
	\includegraphics[scale=0.4]{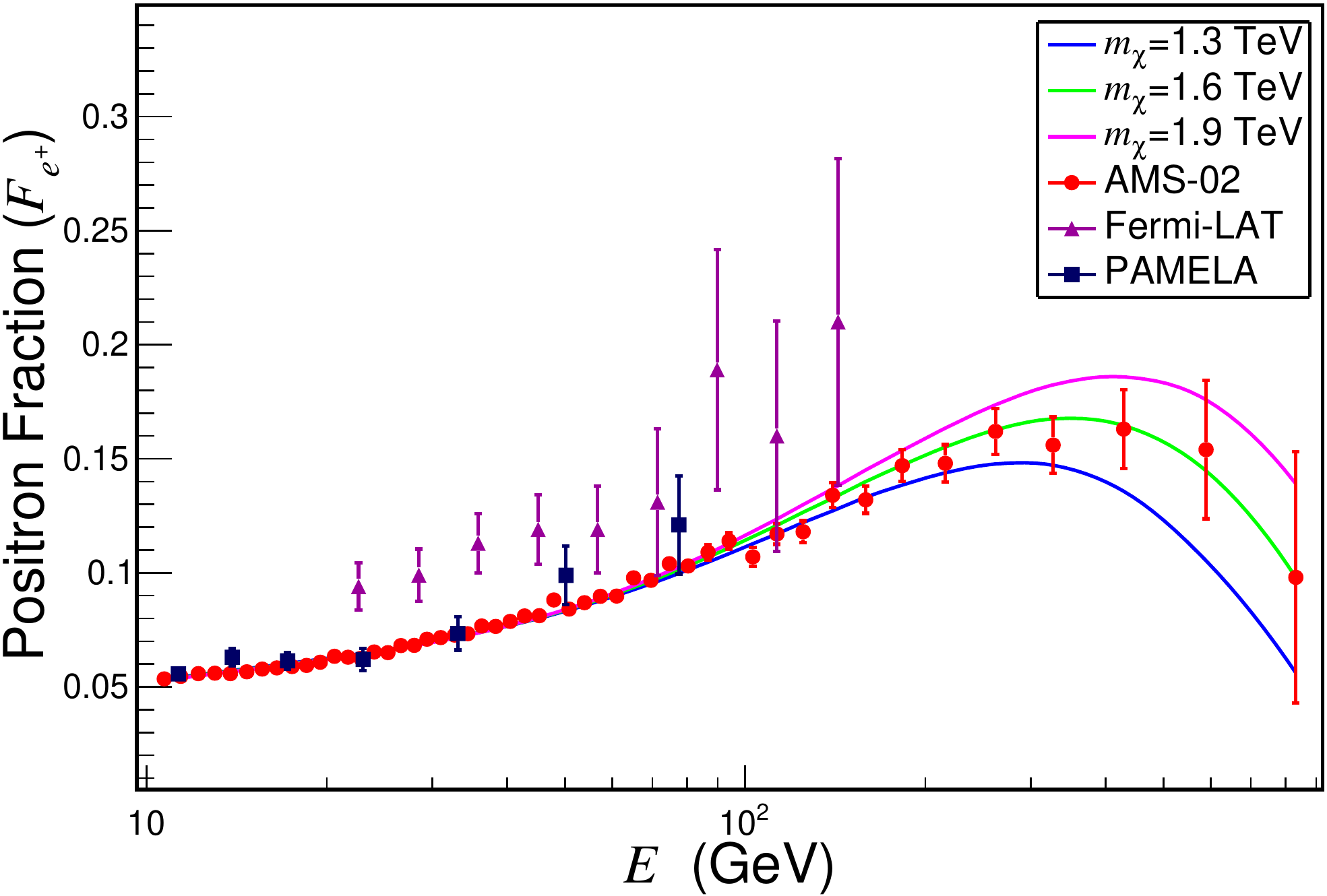}
	\hspace{2 mm}
	\includegraphics[scale=0.4]{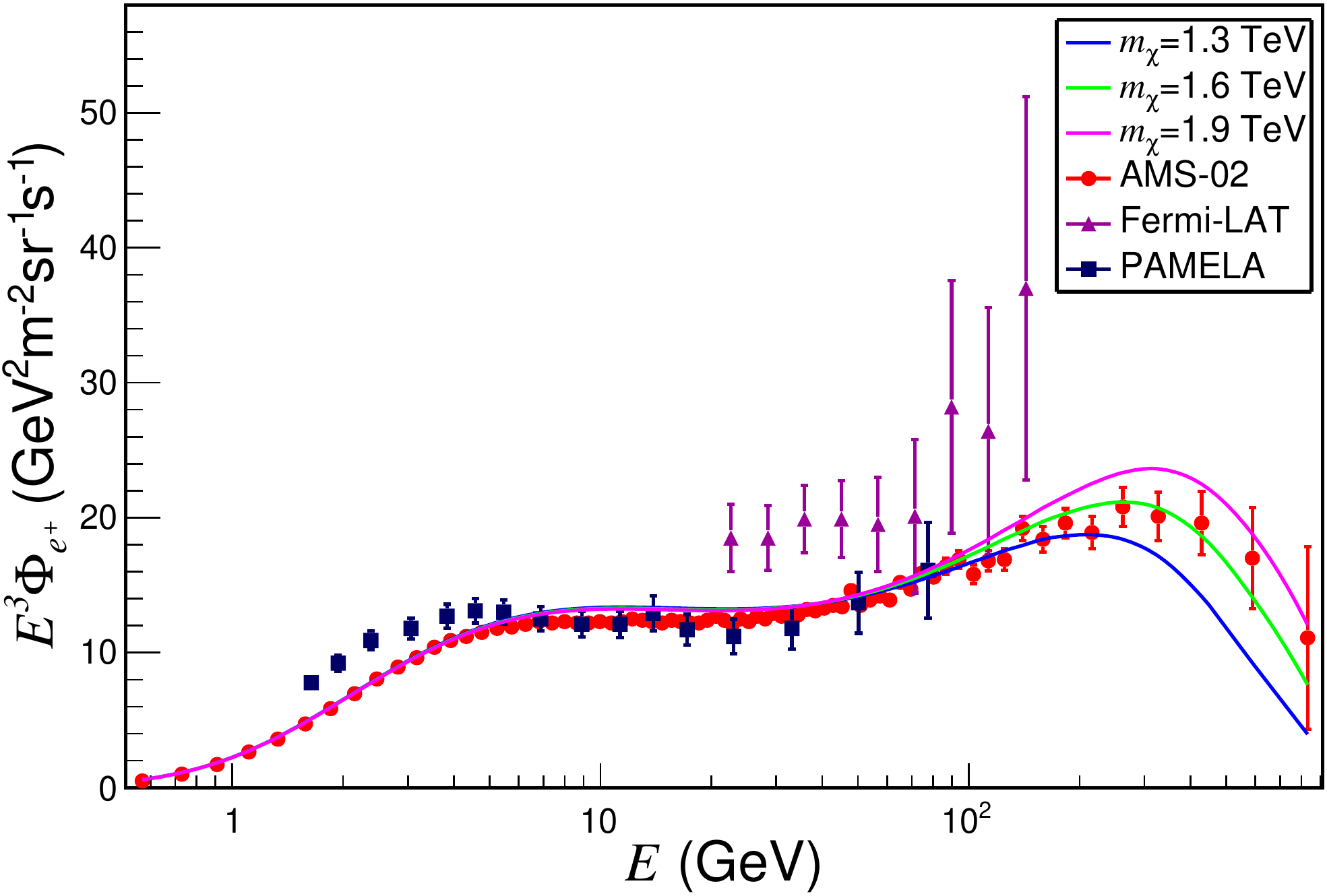}
	\caption{Left Panel: The positron fraction obtained from the DM annihilation for the different benchmark points tabulated in Table \ref{SE_BP}. The experimental data as reported by AMS-02, Fermi-LAT and PAMELA are also shown. Right Panel: The comparison of the positron flux  using the best fit parameters obtained from fitting the AMS-02 observed positron fraction, with the experimental data published by the three experiments.}
	\label{PosFracPlot}
\end{figure}

Following the formalism given above, we compute the positron fraction as obtained from the DM annihilation for the dark matter candidate proposed in this work for various benchmark points. A $\chi^2$ analysis was performed for the data and with the theoretical formalism described in this work. The boost factor $B$ is the free parameter to be evaluated by $\chi^2$ minimization. The best-fit values of the boost factor $B$ (parameter) along with the values of the dark matter mass $m_{\chi}$, the DM-pseudoscalar coupling $y_{\chi}$ and the Sommerfeld enhanced cross-section for the $\bar{\chi}\chi\rightarrow aa$ channel for the different benchmark points are given in Table. \ref{SE_BP}. The final computed positron-fraction results are furnished in Fig. \ref{PosFracPlot} (left panel). Also shown in Fig. \ref{PosFracPlot} (left panel) is the positron fraction data as observed by the AMS-02 ~\cite{AMS_2019_electrons}, Fermi-LAT ~\cite{Fermi-LAT_PosFrac2012} and PAMELA ~\cite{PAMELA_PosFrac2013} experiments. The right panel of Fig. \ref{PosFracPlot} shows the comparison of the positron flux as computed from the present DM model for various benchmark points using the best fit parameters obtained by fitting the positron-fraction data of AMS-02, with that as reported by the AMS-02 ~\cite{AMS_2019_positrons}, Fermi-LAT and PAMELA experiments. For this plot, we consider the diffuse flux to be the same as the positron background model laid out in Eq. (\ref{AMSposBkg}).
It can be seen from Fig.~\ref{PosFracPlot} (left panel) that the dark matter model considered in this work fits well for the excess in positron fraction in the energy region $E\geqslant10$ GeV. Therefore the observational results of AMS-02 as presented in~\cite{AMS_2019_electrons} can be explained by the annihilation of a fermionic dark matter candidate preferably for the dark matter mass of 1.6 TeV through a $\mu$-philic pseudoscalar portal as proposed in this work. It can also be seen that the positron spectrum~\cite{AMS_2019_positrons} could be simultaneously fitted by using the same parameter set as shown in the right panel. We remark that the fit to the positron spectrum is marginally worse than that to the positron fraction.
This discrepancy can be relaxed by choosing an appropriate propagation model of cosmic rays \cite{Lin:2014vja}.

\section{Conclusion}\label{sec:Conclu}
The AMS-02 experiment onboard the International Space Station reported an excess of positron-fraction beyond the positron energy of 10 GeV. While the positron-fraction spectrum below 10 GeV corroborates with the expected spectrum, the excess that peaks around 320 GeV could not be explained by the known astrophysical or other processes, cosmic rays, etc. Therefore, the observed positron-fraction excess could be a signature of new physics or phenomena not fully understood yet. 

In this work, we revisit the possibility that dark matter annihilation in the Universe could have caused this excess signature by producing electron-positron pairs through such annihilation processes. To this end, we formulate a new particle dark matter model based on 2HDM available in the literature and adding to it an extra fermion as a DM candidate and an additional pseudoscalar for the portal. It appears from our calculations that in order to account for the dark matter annihilation cross-section required to satisfy the experimentally observed dark matter relic density given by PLANCK collaboration, a pseudoscalar is needed if the extra fermion in the theory is to serve as the dark matter candidate. It also appears that, for the realization of the observed positron excess within the proposed framework, a leptophilic dark matter could be a viable candidate. Keeping these in view we consider the leptophilic 2HDM which can explain the muon g-2 anomaly while avoiding various experimental and astrophysical constraints as shown in detail. 
The key feature of the model is the presence of a light bosonic portal through which the required DM relic density, as well as the Sommerfeld enhancement of DM annihilation cross-section, can be achieved.
It will be an interesting task to search for such a light boson in various low energy experiments.

We have computed the Sommerfeld enhancement in the present framework and made a $\chi^2$ fit of the computed positron fraction with the AMS-02 observational results where a boost factor is treated as a free parameter. We obtain a good fit for the theoretical estimation of positron-fraction with the boost factor of order 10 which is within the widely accepted ballpark in the literature. We also find that the present dark matter model satisfactorily explains simultaneously the positron spectrum observed by the AMS-02 experiment.

\section*{Acknowledgement}
Three of the authors, SG, ADB and DM, would like to thank P. Roy and A. Ray for some useful discussions. ADB would also like to thank Saha Institute of Nuclear Physics for the hospitality. Work of ADB is supported in part by the National Science Foundation of China (11422545,11947235).
\medskip

\printbibliography

\end{document}